\documentclass[twocolumn,superscriptaddress,nobibnotes,footinbib,floatfix,aps,prb,citeautoscript,reprint]{revtex4-2}

\usepackage{graphicx}
\usepackage{epsfig}
\usepackage{subfigure}
\usepackage{color}
\usepackage{latexsym}
\usepackage{amsmath,bm,multirow}
\usepackage{amsfonts}
\usepackage{dcolumn}           
\usepackage{natbib}
\usepackage[colorlinks,linkcolor=blue,citecolor=green]{hyperref}
\usepackage{physics}

\usepackage[draft]{changes} 
\usepackage[normalem]{ulem}
\definechangesauthor[name={yi}, color=blue]{yi}
\newcommand{\portland}{Department of Mechanical and Materials Engineering, Portland State University, Portland, OR 97201, USA}

\usepackage[draft]{changes} 
\usepackage[normalem]{ulem}
\definechangesauthor[name={Yi}, color=blue]{Yi}
\usepackage{soul}
\begin{document}

\title{Lattice Thermal Transport Beyond the Quasiparticle Approximation: Nontrivial Spectral Competition between Three- and Four-Phonon Interactions}

\author{Yi Xia}
\email{yimaverickxia@gmail.com; yxia@pdx.edu}
\affiliation{\portland}

\date{\today}

\begin{abstract}

The breakdown of the quasiparticle approximation (QPA) for phonons in strongly anharmonic materials necessitates advanced first-principles frameworks for accurate lattice dynamics and thermal transport predictions. We develop a comprehensive beyond-quasiparticle approximation (BQPA) approach incorporating both three-  (3ph) and four-phonon (4ph) interactions and apply it to investigate lattice thermal conductivity ($\kappa_{\rm L}$) in MgO, PbTe, and AgCl -- materials that span a broad spectrum of anharmonicity, from weak to severe anharmonic regimes with overdamped phonons. We reveal that while BQPA consistently increases $\kappa_{\rm L}$ relative to QPA due to phonon softening when considering only 3ph interactions, the inclusion of additional 4ph interactions hardens the phonon spectrum and suppresses this enhancement, bringing BQPA and QPA predictions into close agreement via subtle spectral competition effects across all three compounds. These findings highlight that accurate modeling of $\kappa_{\rm L}$ in strongly anharmonic materials requires treating both full phonon spectral function and higher-order anharmonicity on equal footing. Our work establishes a systematic framework for modeling thermal transport in systems with overdamped phonons and provides critical insights for materials design beyond the limits of conventional phonon transport theory.

\end{abstract}

\maketitle

Lattice thermal conductivity ($\kappa_{\rm L}$), defined as the ability of a material to transfer heat under a temperature gradient via atomic vibrations~\cite{nellis2008heat}, is a fundamental property of materials and plays critical role across diverse technologies, including microelectronics~\cite{shinde2006high,he2021thermal} , thermal barrier coatings~~\cite{suresh1997investigation,vassen2000zirconates}, and thermoelectrics~\cite{bell2008cooling,he2017advances}. The drive to discover materials with extreme $\kappa_{\rm L}$, from ultra-high (e.g., BAs~\cite{Lindsay2013,Kang575,Tian582,Li579,NIYIKIZA2025} and BN~\cite{chen2020ultrahigh}) to ultra-low (e.g., SnSe~\cite{zhao2014ultralow} and Tl$_3$VSe$_4$~\cite{mukhopadhyay2018two}), has led to significant advances in thermal transport science. Moreover, emerging quantum materials~\cite{keimer2017physics,kim2023chiral}, where heat can be carried and manipulated by quantum excitations such as chiral phonons~\cite{zhang2015chiral,ishito2023truly,kim2023chiral,wang2024chiral,ueda2023chiral} and topological electrons~\cite{qi2011topological}, offer new opportunities to control heat flow at the nanoscale, with potential applications in spintronics and quantum information processing~\cite{awschalom2002semiconductor}. To fully harness the potential of these materials in their respective applications, a fundamental understanding of the microscopic mechanisms governing $\kappa_{\rm L}$ is essential for enhancing material performance and enabling the rational design of materials with tailored thermal properties.

Recent years have seen significant advances in the simulation of $\kappa_{\rm L}$ within and beyond the first-principles framework that combines anharmonic lattice dynamics (ALD) and Peierls-Boltzmann transport equation (PBTE)~\cite{broido2007intrinsic}. 
These advances can be categorized into two main aspects. The first aspect has been focused on an improved description of phonon quasiparticles using many-body theory for multi-phonon interactions. For instance, self-consistent phonon (SCPH) theory~\cite{SCPH2015,tadano2022first}, temperature-dependent effective potential (TDEP)~\cite{TDEP2011}, and the stochastic self-consistent harmonic approximation (SSCHA)~\cite{errea2014anharmonic,bianco2017second,SSCHA}, along with its dynamical extension~\cite{monacelli2021time,lihm2021gaussian}, have been developed to better describe vibrational frequencies at finite temperatures. Additionally, the quantum mechanical formalisms of four-phonon (4ph) scatterings beyond three-phonon (3ph) picture have been developed and implemented~\cite{Tianli2016, Tianli2017} to account for anharmonicity-induced phonon broadening. Recent studies have demonstrated that both phonon frequency shifts and broadening due to higher-order anharmonicity play a crucial role in impacting $\kappa_{\rm L}$ in common semiconductors~\cite{rczbprx, ravichandran2018unified}. The second aspect of the advance involves generalizing the PBTE into a unified theory of thermal transport in crystals and glasses~\cite{Simoncelli2019}. This unified theory has been developed using either the Wigner formulation~\cite{Simoncelli2022} or Green-Kubo formalism~\cite{Isaeva:2019aa}. The unified theory has enabled a reliable computation of $\kappa_{\rm L}$ in complex crystals, thus bridging the gap in heat transfer between the particle-like phonon propagation and the wave-like tunneling~\cite{Simoncelli2019}.

Despite these encouraging advances, existing studies on $\kappa_{\rm L}$ combining ALD and PBTE or Wigner formulation have predominantly relied on the quasiparticle approximation (QPA)~\cite{kaxiras2003atomic}. When applied to the phonons, the validity of QPA requires that the relative strength of the phonon linewidth ($\Gamma$) with respect to the vibrational frequency ($\omega$) satisfies $\Gamma \ll \omega$. While QPA is successful in many cases, it fails to fully capture physics underlying the complex vibrational spectrum in materials with strong phonon scattering. Specific examples include perovskites with dynamical instabilities, such as CsPbBr$_3$, where two-dimensional overdamped phonons are experimentally observed~\cite{CsPbBr32021}; thermoelectrics with low $\kappa_{\rm L}$, exemplified by the presence of multiple quasiparticle peaks in PbTe~\cite{PbTe2011} and SnSe~\cite{aseginolaza2019phonon}; and Earth's mantle materials like MgO, where phonon band theory has been shown to break down at extreme temperatures~\cite{coiana2024breakdown}.

In recognizing the breakdown of QPA in these scenarios, efforts on going beyond the quasiparticle approximation (BQPA) have been attempted~\cite{xie2022dynamics,dangic2021origin,dangic2025lattice}. To account for the non-Lorentzian phonon lineshapes, a Green-Kubo approach in combination with TDEP was implemented and applied to investigate $\kappa_{\rm L}$ of GeTe near the phase transition~\cite{dangic2021origin}. The BQPA effects were accounted for by means of full phonon spectral function (FPSF). Most recently, the mentioned Green-Kubo linear response theory was further integrated with the SSCHA, which was leveraged to study thermal transport in CsPbBr$_{3}$ across structural phase transitions~\cite{dangic2025lattice}. Both studies~\cite{dangic2021origin,dangic2025lattice} reported consistent increases in $\kappa_{\rm L}$ (10\% - 25\%) from BQPA compared to QPA. This observation is intriguing and raises an important question: does BQPA inherently predict larger values of $\kappa_{\rm L}$ compared to QPA? This question is highly relevant and important for the quantitative predictions of $\kappa_{\rm L}$ for materials with strong anharmonicity~\cite{pal2019high,zeng2021ultralow,ouyang2023role,xie2020first}.

Furthermore, since both studies~\cite{dangic2021origin,dangic2025lattice} relied solely on 3ph interactions (cubic anharmonicity) for the dynamical response of phonons to construct FPSFs, critical questions remain in the regime of BQPA regarding the role of 4ph scatterings (quartic anharmonicity), which have been shown to be significant in many materials~\cite{Tianli2017, ravichandran2018unified, pbte2018, xiao2023anharmonic, yang2019stronger}, even at room temperature~\cite{rczbprx}. In this Letter, we address these questions through three main objectives: (1) to develop a comprehensive microscopic lattice dynamics model that incorporates both cubic and quartic anharmonic interactions for computing FPSFs; (2) to integrate this model with advanced thermal transport theory to explicitly compute $\kappa_{\rm L}$ in the BQPA regime; and (3) to elucidate the distinctive contributions of 3ph and 4ph interactions to $\kappa_{\rm L}$.

\begin{figure}[htp]
	\includegraphics[width = 0.95\linewidth]{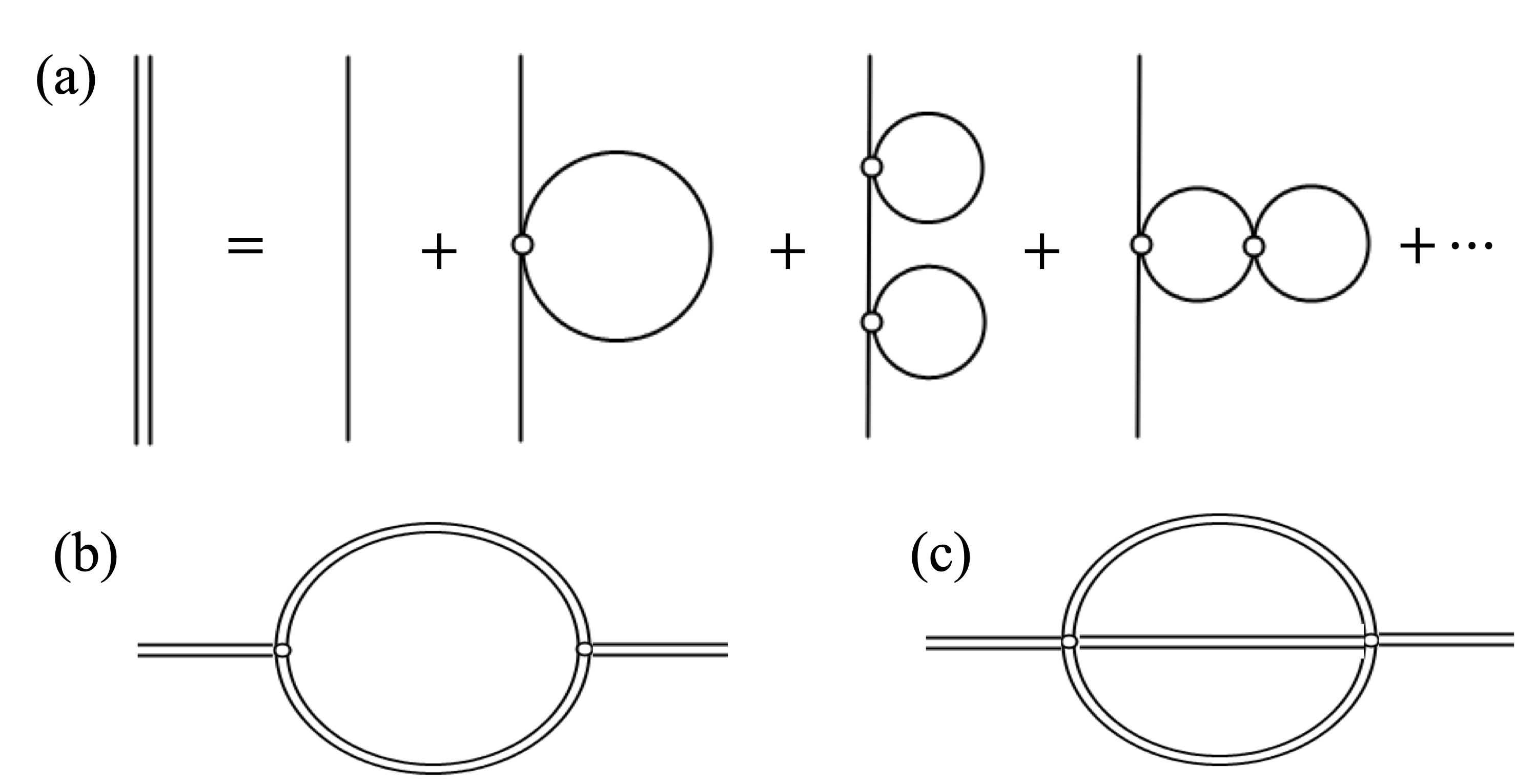}
	\caption{ 
	(a) Diagrammatic representation of the self-consistent phonon (SCPH) propagator where the loop diagram is considered (also termed as Hartree-Fock approximation for phonons~\cite{xiao2023anharmonic}). The double line represents the SCPH propagator, while the sold line is that of a free phonon. (b) Bubble (3ph interactions) and (c) sunset (4ph interactions) diagrams defined on the basis of the SCPH propagator.
	}
	\label{fig:diagram}
\end{figure}


\begin{figure*}[htp]
	\includegraphics[width = 1.0\linewidth]{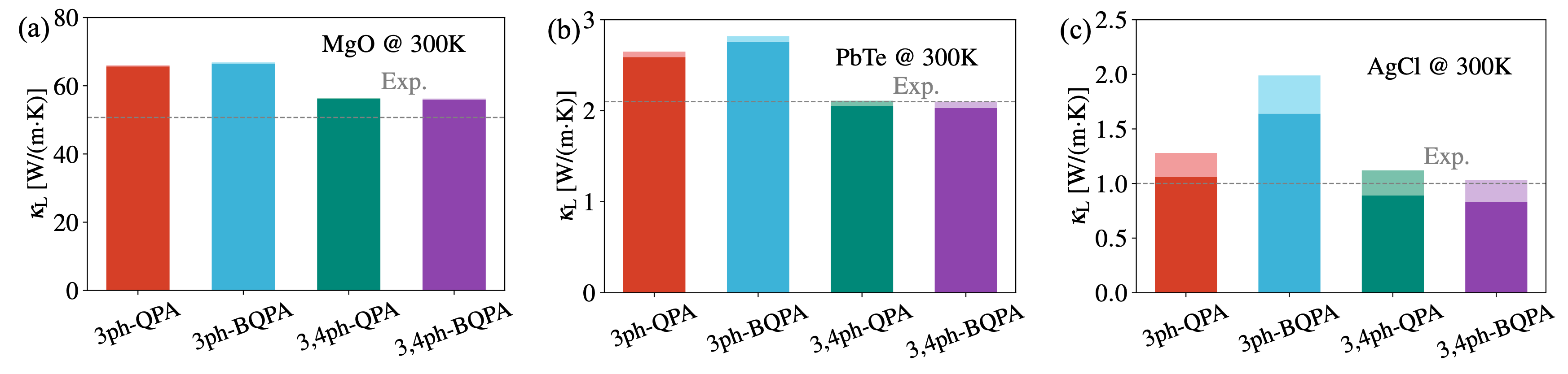}
	\caption{ 
	Lattice thermal conductivity comparison across four theoretical approaches: 3ph-QPA (three-phonon interactions within the quasiparticle approximation), 3ph-BQPA (three-phonon interactions beyond the quasiparticle approximation), 3,4ph-QPA (three- and four-phonon interactions within the quasiparticle approximation), and 3,4ph-BQPA (three- and four-phonon interactions beyond the quasiparticle approximation) for (a) MgO, (b) PbTe, and (c) AgCl at 300 K. Self-consistent phonon is implicitly implied in all adopted theoretical approaches. Experimental values~\cite{cahill1998thermal,el1983thermophysical,maqsood2004thermophysical} at 300~K are shown as horizontal gray dashed lines for comparison. Dark and light portions of the bars represent the particle-like and wave-like/coherent contributions to the lattice thermal conductivity, respectively.
	}
	\label{fig:kappa}
\end{figure*}



We start by briefly describing a theoretical framework relying on many-body Green's function approach to model $\kappa_{\rm L}$, recently developed by Caldarelli \textit{et al.}~\cite{caldarelli2022many}. Their approach features the adoption of Wigner heat flux with an explicit inclusion of coherent thermal transport~\cite{Simoncelli2022} and the derivation of $\kappa_{\rm L}$ using finite-temperature Green's functions, the latter of which naturally extend to the regime where QPA breaks down. The resulting formula for calculating $\kappa_{\rm L}$ in the regime of BQPA reads
\begin{equation}\label{eq:kappa_bqpa}
\begin{split}
    \kappa^{\alpha\beta}_{\rm BQPA}  = &   \frac{\hbar^2\pi}{N_c V k_{\rm B} T^2} 
    \sum_{\mathbf{q}, ss^{\prime}}
    \frac{ ( \omega_{\mathbf{q},s} + \omega_{\mathbf{q},s^{\prime}} )^2 }{4}
    v^{\alpha}_{\mathbf{q},{ss^{\prime}}}
    v^{\beta}_{\mathbf{q},{s^{\prime}s}} \\
    & \cdot \int d\omega b_{\mathbf{q},s^{\prime}}(\omega) 
    b_{\mathbf{q},s}(\omega)
    n(\omega)(n(\omega)+1),
\end{split}
\end{equation}

\noindent where $\hbar$, $k_{\rm B}$, $N_c$, $V$, and $T$ are, respectively, the reduced Planck constant, Boltzmann's constant, the number of sampled wave vectors, the volume of the primitive cell, and the absolute temperature. The phonon mode-resolved properties, namely, $\omega_{\mathbf{q},s}$ and $v^{\alpha}_{\mathbf{q},{ss^{\prime}}}$ are the frequency and generalized group velocity (see Eq.[34] in Ref.\cite{caldarelli2022many}) for mode with the wave vector $\mathbf{q}$ and branch $s$, and $\alpha/\beta$ is a Cartesian index. All of these quantities can be readily obtained from harmonic phonons. The key to the beyond-quasiparticle description lies in the integral part of Eq.(\ref{eq:kappa_bqpa}), wherein $n(\omega)$ is the frequency-dependent phonon population that obeys the Bose-Einstein statistics, and $b_{\mathbf{q}, s}(\omega)$ is the frequency-dependent phonon spectral density~\cite{caldarelli2022many}
\begin{equation}\label{eq:phonon_spectral}
    b_{\mathbf{q},s}(\omega) = \frac{1}{\pi} 
    \frac{\gamma_{\mathbf{q}, s}(\omega)}{(\omega - \omega_{\mathbf{q},s} - \Delta_{\mathbf{q},s}(\omega) )^2 + \gamma_{\mathbf{q},s}(\omega)^2},
\end{equation}
where the frequency-dependent phonon self-energy is defined as $\Sigma_{\mathbf{q},s}(\omega) = \hbar \Delta_{\mathbf{q},s}(\omega) -i \hbar \gamma_{\mathbf{q},s}(\omega)$, with $\Delta$ and $\gamma$ denoting the frequency-dependent real and imaginary part of phonon self-energy, respectively.

We see from Eq.(\ref{eq:phonon_spectral}) that an accurate calculation of $\kappa_{\rm BQPA}$ depends on a precise estimation of $\Sigma_{\mathbf{q},s}(\omega)$. In this work, we employed the first-order self-consistent phonon (SCPH) approach, truncating the self-energy to the loop diagram [Fig.~\ref{fig:diagram}(a)], to estimate static phonon frequencies. As discussed in Ref.[\onlinecite{xiao2023anharmonic}] and [\onlinecite{monacelli2025analyzing}], this method can be interpreted as the Hartree-Fock (HF) approximation for interacting phonons. Building on SCPH, we then computed frequency-dependent $\Sigma_{\mathbf{q},s}(\omega)$ by incorporating contributions from both bubble (3ph interactions) [Fig.~\ref{fig:diagram}(b)] and sunset (4ph interactions) [Fig.~\ref{fig:diagram}(c)] diagrams, i.e., $\Sigma_{\mathbf{q},s}(\omega) \approx \Sigma_{\mathbf{q},s}^{3ph}(\omega) + \Sigma_{\mathbf{q},s}^{4ph}(\omega)$ (see Appendix~\ref{sec:selfeng} for detailed expressions). To avoid double-counting when evaluating $\Sigma_{\mathbf{q},s}^{3ph}(\omega)$, we chose not to include the real part of the bubble diagram in the SCPH calculation, as opposed to Ref.[\onlinecite{tadano2022first}]. We note that the above theoretical framework is grounded in rigorous many-body diagrammatic perturbation theory~\cite{xiao2023anharmonic}, and, as correctly emphasized in Ref.[\onlinecite{monacelli2025analyzing}], it should not be conflated with the TDEP~\cite{TDEP2011} or SSCHA~\cite{SSCHA} approaches.

Having established the theoretical framework, we now turn to explore the anharmonic lattice dynamics and thermal transport properties of materials with varying degrees of anharmonicity, spanning from the QPA to BQPA. For this investigation, we selected three face-centered cubic materials: MgO, PbTe, and AgCl. MgO exhibits relatively weak anharmonicity and plays a significant role in Earth's mantle materials. PbTe, in contrast, demonstrates stronger anharmonicity, which is essential for thermoelectric applications. Finally, AgCl displays severe anharmonicity, with overdamped phonon vibrations ($\Gamma > \omega$) at room temperature~\cite{rczbprx}. To evaluate Eq.~(\ref{eq:phonon_spectral}), we utilize the harmonic and anharmonic force constants reported in the literature: MgO from Ref.[\onlinecite{xia2025first}], PbTe from Ref.[\onlinecite{pbte2018}], and AgCl from Ref.[\onlinecite{rczbprx}] (see Appendix~\ref{sec:details} for computational details). For all three materials, both 3ph and 4ph interactions were computed using a uniform phonon wave vector mesh of 25$\times$25$\times$25 sampling points in the first Brillouin zone, which has been tested to achieve good convergence of both Eq.[\ref{eq:kappa_bqpa}] and [\ref{eq:phonon_spectral}].


\begin{figure*}[htp]
	\includegraphics[width = 1.0\linewidth]{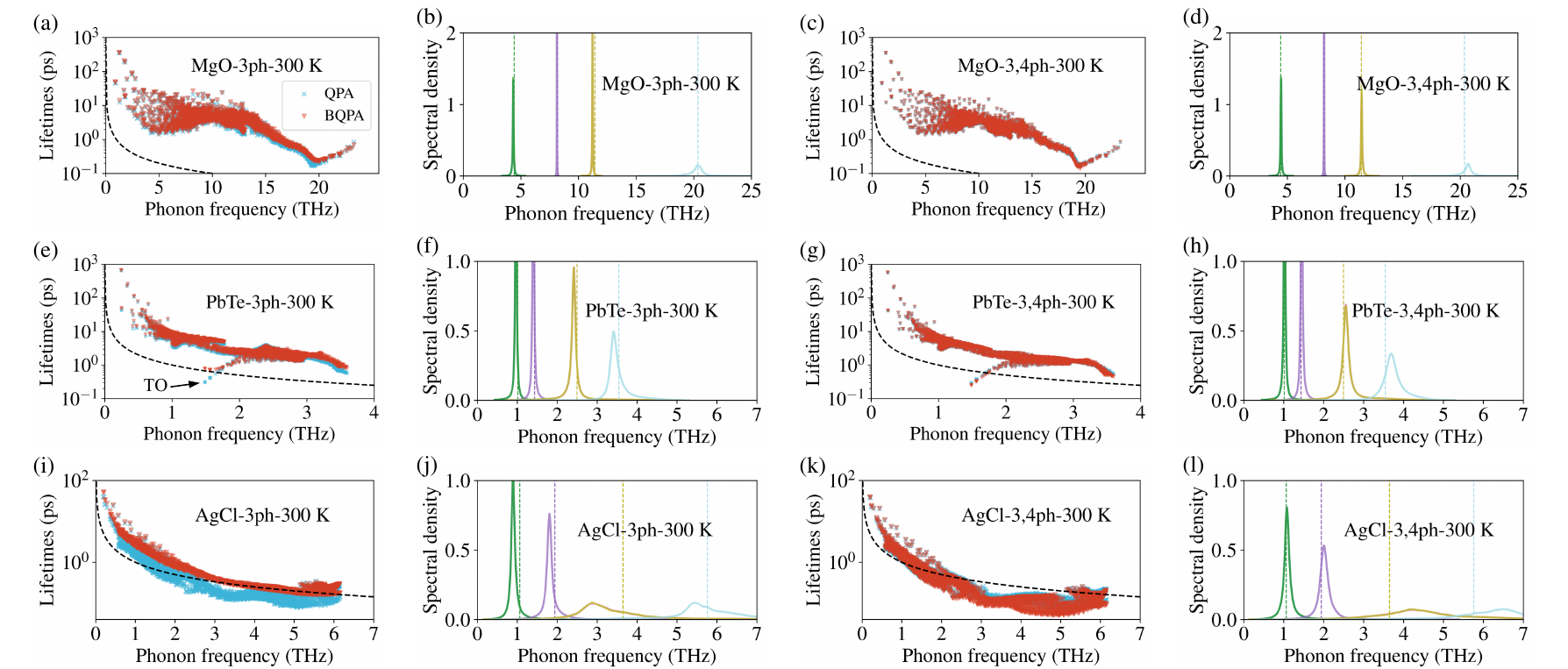}
	\caption{ 
	(a) Comparison of calculated phonon lifetimes for three-phonon (3ph) scatterings within the quasiparticle approximation (QPA) and beyond the quasiparticle approximation (BQPA) from Eq.(\ref{eq:renorm_lifetime}) for MgO at 300~K. (b) Mode-resolved phonon spectral density (solid lines) considering only 3ph interactions compared with corresponding self-consistent phonon frequencies (vertical dashed lines) for MgO at 300~K. The selected phonon mode is at (0.2, 0, 0) along the $\Gamma$ to $X$ path. (c,d) Similar analysis to (a,b) but including additional four-phonon (4ph) interactions. (e-h) Corresponding analysis for PbTe at 300 K following the same theoretical approaches as panels (a-d). (i-l) Similar analysis for AgCl at 300~K. The black dashed lines in (a), (c), (e), (g), (i), and (k) indicate lifetimes equal to the inverse frequency. Full phonon spectral density for MgO, PbTe, and AgCl at 300~K along the $\Gamma-X$ direction in the first Brillouin zone is shown in Fig.~\ref{fig:em_fpsf} in Appendix~\ref{sec:details}.
	}
	\label{fig:rates}
\end{figure*}


To better reveal the effects of 4ph interactions and BQPA, we compute $\kappa_{\rm L}$ using four different theoretical approaches: 3-phonon interactions within the quasiparticle approximation (3ph-QPA), 3- and 4-phonon interactions within QPA (3,4ph-QPA), 3-phonon interactions within the beyond quasiparticle approximation (3ph-BQPA), and 3- and 4-phonon interactions within BQPA (3,4ph-BQPA). We note that SCPH is implicitly implied, and that QPA refers to the adoption of Lorentz approximation for $b_{\mathbf{q},s}(\omega)$, where the phonon lifetime ($\tau_{\mathbf{q},s}$) is given by the inverse of the scattering rate (see Eq.(\ref{eq:kappa_qpa}) in Appendix~\ref{sec:quasi}). The calculated $\kappa_{\rm L}$ using these four levels of theories for MgO, PbTe, and AgCl at 300~K is displayed in Fig.~\ref{fig:kappa}.

For MgO, considering only 3ph interactions, BQPA yields a slight increase in $\kappa_{\rm L}$ compared to QPA. However, when both 3ph and 4ph interactions are included, the difference between BQPA and QPA becomes minimal. The inclusion of 4ph interactions reduces $\kappa_{\rm L}$ by approximately 10\% at 300 K, consistent with literature reports~\cite{kwon2020dominant,han2023predictions}, and both 3,4ph-QPA and 3,4ph-BQPA approaches show good agreement with experimental measurements. This behavior is expected for MgO, as QPA should be a good approximation at room temperature due to the material's relatively weak anharmonicity.

For PbTe and AgCl, different trends emerge that highlight the role of anharmonicity strength. In PbTe, BQPA increases $\kappa_{\rm L}$ by about 7\% compared to QPA when considering only 3ph interactions, while the inclusion of 4ph interactions (which reduces $\kappa_{\rm L}$ by ~21\% at 300~K under QPA) brings both methods into close agreement, with BQPA showing a slightly enhanced coherent contribution. For the strongly anharmonic AgCl, BQPA produces a dramatic increase in both particle-like (55\%) and coherent (59\%) contributions to $\kappa_{\rm L}$ under 3ph-only calculations, but this enhancement in both particle-like and coherent channels is largely suppressed when 4ph interactions are included, yielding results similar between QPA and BQPA and achieving reasonable experimental agreement.

Remarkably, across all three compounds, BQPA consistently increases $\kappa_{\rm L}$ when only 3ph interactions are considered, but the inclusion of 4ph interactions tends to bring QPA and BQPA results into alignment. These observed trends naturally raise the question: why do 3ph and 4ph interactions have such different effects within BQPA?
To answer these question, we trace back to Eq.(\ref{eq:kappa_bqpa}) and compare it with its form under QPA [Eq.(\ref{eq:kappa_qpa}) in Appendix~\ref{sec:quasi}]. To enable a more straightforward comparison, we define the lifetime under BQPA as
\begin{equation}\label{eq:renorm_lifetime}
    \tau_{\mathbf{q},ss^{\prime}}^{\rm BQPA} =\pi
    \frac{\int d\omega b_{\mathbf{q},s^{\prime}}(\omega) 
    b_{\mathbf{q},s}(\omega)
    n(\omega)(n(\omega)+1)}
    {n_{\mathbf{q},s}(n_{\mathbf{q},s^{\prime}}+1)}.
\end{equation}
By noticing that phonon mode heat capacity has a form of  $C_{\mathbf{q},s}=\hbar^2\omega_{\mathbf{q},s}^2 n_{\mathbf{q},s}(n_{\mathbf{q},s}+1)/(k_{\rm B}T^2)$ under QPA, we can rewrite Eq.(\ref{eq:kappa_bqpa}) as $\kappa_{\rm BQPA}^{\alpha\beta} = 1/(N_cV)\sum_{\mathbf{q},s=s^{\prime}} v^{\alpha}_{\mathbf{q}, ss^{\prime}}
v^{\beta}_{\mathbf{q},s^{\prime}s} C_{\mathbf{q},s}\tau_{\mathbf{q},ss^{\prime}}^{\rm BQPA}$ in the absence of coherent contribution (i.e., summation is carried over $s=s^{\prime}$). With this form, it is clear that the difference in BQPA and QPA comes solely from the difference between $\tau_{\mathbf{q},s}^{\rm BQPA}$ (equivalent to $\tau_{\mathbf{q},ss^{\prime}}^{\rm BQPA}$ with $s=s^{\prime}$) and $\tau_{\mathbf{q},s}^{\rm QPA}$.

In Fig.~\ref{fig:rates}, we compare $\tau_{\mathbf{q},s}^{\rm BQPA}$ and $\tau_{\mathbf{q},s}^{\rm QPA}$ for all three compounds, considering 3ph and 4ph interactions step by step. For MgO, both lifetimes show similar values with or without 4ph interactions [Fig.~\ref{fig:rates}(a) and (c)], though $\tau_{\mathbf{q},s}^{\rm BQPA}$ is slightly enhanced in the 3ph-only case, consistent with its weak anharmonicity. This is further supported by the phonon spectral density [Fig.~\ref{fig:rates}(b) and (d)], which exhibits sharp Lorentzian peaks near SCPH frequencies, confirming the validity of QPA. In PbTe, $\tau_{\mathbf{q},s}^{\rm BQPA}$ and $\tau_{\mathbf{q},s}^{\rm QPA}$ remain similar in both 3ph and 3,4ph cases, except for transverse optical modes near the Brillouin zone center, where BQPA lifetimes are strongly enhanced due to softening of the phonon spectral density peak relative to the SCPH frequencies [Fig.~\ref{fig:rates}(f)]. It is because, as per Eq.~(\ref{eq:renorm_lifetime}), such softening leads to a larger weight of $n(\omega)(n(\omega)+1)$ in the low-frequency end compared with $n_{\mathbf{q},s}(n_{\mathbf{q},s}+1)$, thus leading to enhanced $\tau_{\mathbf{q},s}^{\rm BQPA}$. We note that similar trend has also been observed in Ref.[\onlinecite{dangic2021origin}] for GeTe under the consideration of only 3ph interactions. However, this effect vanishes when spectral density is hardened by 4ph interactions [Fig.~\ref{fig:rates}(g)]. A similar behavior is observed in AgCl, where strong softening under 3ph leads to significant lifetime enhancement under BQPA [Fig.~\ref{fig:rates}(i)], resulting in much larger $\kappa_{\rm L}$ from 3ph-BQPA compared to 3ph-QPA, as discussed earlier. Including 4ph interactions hardens the spectral density [Fig.~\ref{fig:rates}(l)], which mitigates the lifetime enhancement and even slightly decreases lifetime [Fig.~\ref{fig:rates}(k)]. Therefore, we see that the similar values between 3,4ph-QPA and 3,4ph-BQPA have a nontrivial physical origin arising from subtle phonon spectral competition effects. Given that phonon hardening due to 4ph interactions has also been observed in AgCrSe$_2$~\cite{xie2020first}, similar effects may also occur in other materials, highlighting the need for a systematic investigation across various material types.

\begin{figure}[htp]
	\includegraphics[width = 0.85\linewidth]{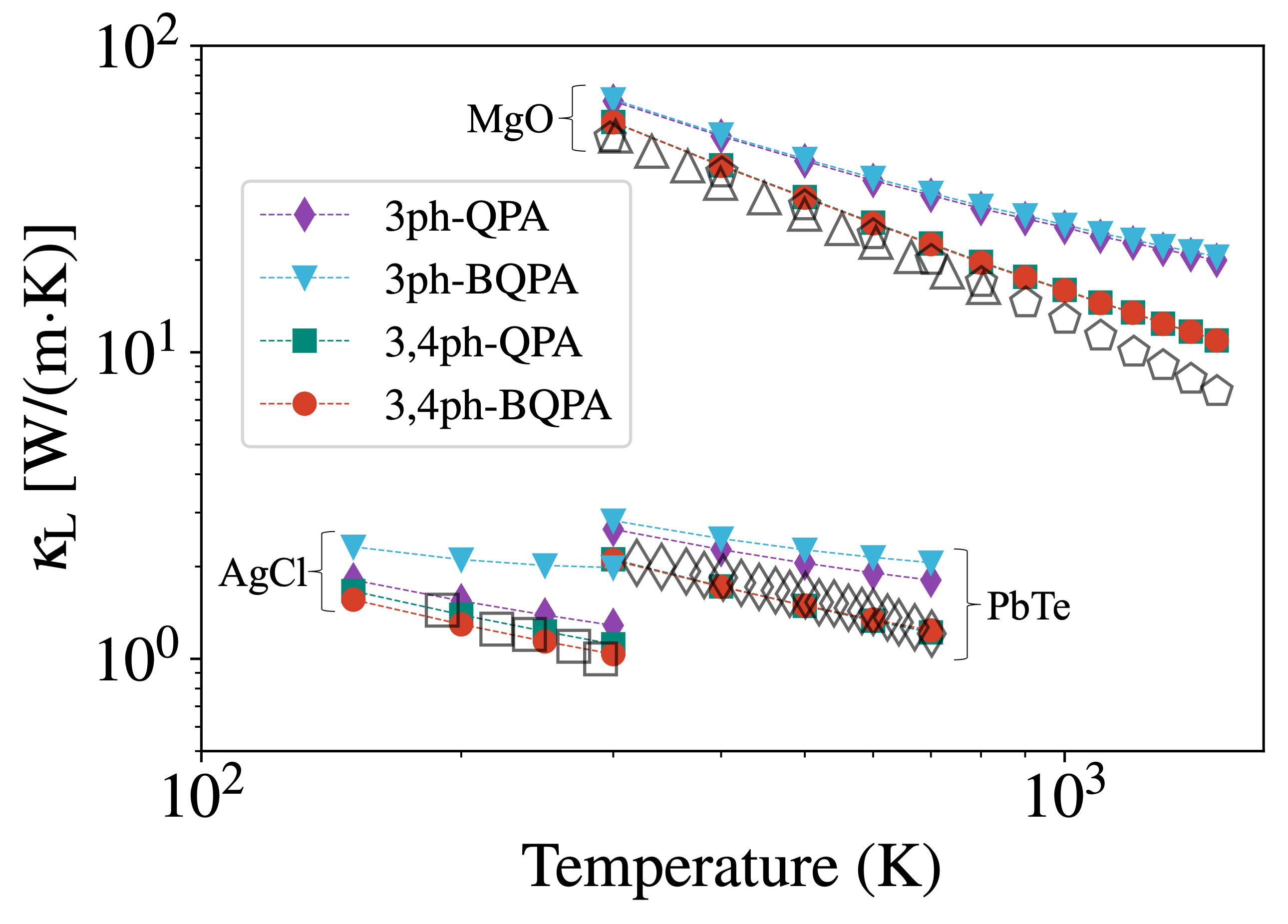}
	\caption{
	Calculated temperature-dependent lattice thermal conductivities using four theoretical approaches (3ph-QPA, 3ph-BQPA, 3,4ph-QPA, and 3,4ph-BQPA) in comparison with experimental measurements, which are shown as empty symbols 
    (squares from Ref.~[\onlinecite{maqsood2004thermophysical}]; diamonds from Ref.~[\onlinecite{el1983thermophysical}]; triangles from Ref.~[\onlinecite{cahill1998thermal}]; pentagons from Ref.~[\onlinecite{hofmeister2014thermal}]).
	}
	\label{fig:Tkappa}
\end{figure}

In Fig.~\ref{fig:Tkappa}, we demonstrate the temperature-dependent $\kappa_{\rm L}$ calculated using four theoretical approaches compared against experimental measurements. Considering only 3ph interactions, the disparity between 3ph-QPA and 3ph-BQPA increases with both material anharmonicity (from MgO to AgCl) and temperature, highlighting the 3ph-induced softening of phonon spectral functions. With additional 4ph interactions, both 3,4ph-QPA and 3,4ph-BQPA yield consistently lower $\kappa_{\rm L}$ compared to 3ph-BQPA, while the hardening of phonon spectral functions induced by 4ph interactions approximately equalizes the predictions between these two methods, except in severely anharmonic AgCl where 3,4ph-BQPA is slightly lower than 3,4ph-QPA. Overall, reasonable agreement with experiment is achieved by both 3,4ph-QPA and 3,4ph-BQPA approaches, with BQPA performing slightly better. Therefore, our results highlight the importance of including both four-phonon interactions and full phonon spectral functions to accurately investigate the underlying microscopic mechanisms and to capture $\kappa_{\rm L}$ across a wide range of anharmonic materials. It is anticipated that strong impacts on $\kappa_{\rm L}$ due to beyond-quasiparticle effects may arise when the combined effects of 3ph and 4ph interactions lead to an overall pronounced phonon softening or hardening.

We have now fully accounted for the spectral effects on $\kappa_{\rm L}$ via Eq.~(\ref{eq:phonon_spectral}); however, another important factor, the renormalization of phonon group velocities due to phonon softening or hardening, remains unaddressed. To assess its influence, we evaluate how the phonon softening induced by 3ph interactions in AgCl at 300 K affects the resulting $\kappa_{\rm L}^{\rm BQPA}$. As shown in Fig.~\ref{fig:velrenorm} and discussed in Appendix~\ref{sec:group}, we find that group-velocity renormalization reduces $\kappa_{\rm L}^{\rm BQPA}$, consistent with the observed phonon softening, but its impact is weaker than that of the spectral effects. In the presence of additional 4ph interactions, we expect an analogous trend, with phonon hardening leading to a slight increase in $\kappa_{\rm L}^{\rm BQPA}$, a direction that could be explored more thoroughly in future work.


To summarize, the beyond-quasiparticle framework incorporating higher-order anharmonicity developed in this Letter provides a systematic computational tool for analyzing lattice dynamics and thermal transport in materials characterized by severe anharmonicity and overdamped phonons. While our focus has been on simple rocksalt structures, the framework can be readily extended to other complex compounds, with perovskites being of particular interest due to their technological importance and rich anharmonic behavior. The systematic diagrammatic treatment of phonon spectral density established here opens possibilities for future investigations that include even higher-order anharmonicity, such as five- and six-phonon interactions recently developed by us~\cite{xia2025first}, when necessary for capturing the full complexity of strongly anharmonic systems.

\begin{acknowledgments}
\textbf{Acknowledgments:} 
Y. X. acknowledges 1) the support from the US National Science Foundation through award CBET-2445361, 2) the support from the Faculty Development Program at Portland State University, and 3) the computing resources provided by Bridges2 at Pittsburgh Supercomputing Center (PSC) through allocations mat220006p and mat220008p from the Advanced Cyber-infrastructure Coordination Ecosystem: Services \& Support (ACCESS) program, which is supported by National Science Foundation grants 2138259, 2138286, 2138307, 2137603, and 2138296. Y. X. is grateful to Z.J. W. for her encouragement and support during the preparation of this manuscript.
\end{acknowledgments}

\bibliography{CuSbS}

\appendix


\section{Quasiparticle approximation}\label{sec:quasi}

As shown in Ref.~[\onlinecite{caldarelli2022many}], when the Lorentzian spectral function approximation (LSFA) is adopted -- i.e., $\gamma_{\mathbf{q},s} \ll \omega_{\mathbf{q},s}$, where the quasiparticle approximation is valid -- Eq.~(\ref{eq:kappa_bqpa}) reduces to the following expression:

\begin{equation} \label{eq:kappa_qpa}
\begin{split}
    \kappa^{\alpha\beta}_{\rm QPA} = & \frac{1}{N_c V} 
    \sum_{\mathbf{q}, ss^{\prime}}
    v^{\alpha}_{\mathbf{q}, ss^{\prime}}
    v^{\beta}_{\mathbf{q},s^{\prime}s}
    \left[  \frac{C_{\mathbf{q},s}}{\omega_{\mathbf{q},s}} +
    \frac{C_{\mathbf{q},s^{\prime}}}{\omega_{\mathbf{q},s^{\prime}}}
    \right] \\
    & \cdot \frac{(\Gamma_{\mathbf{q},s}+ \Gamma_{\mathbf{q},s^{\prime}}) (\omega_{\mathbf{q},s} + \omega_{\mathbf{q},s^{\prime}}) }
    { 8(\omega_{\mathbf{q},s} - \omega_{\mathbf{q},s^{\prime}} )^2  + 2(\Gamma_{\mathbf{q},s}+ \Gamma_{\mathbf{q},s^{\prime}})^2},
\end{split}
\end{equation}

\noindent where $C_{\mathbf{q},s}$ is the modal heat capacity and $\Gamma_{\mathbf{q},s} \equiv 2\gamma_{\mathbf{q},s} $ is the scattering rate. The above expression is also consistent with the Wigner transport equation (WTE)~\cite{Simoncelli2019,Simoncelli2022}. Moreover, in the absence of coherent transport, Eq.(\ref{eq:kappa_qpa}) is further reduced to 
\begin{equation}
\kappa_{\rm QPA}^{\alpha\beta} = \frac{1}{N_cV}\sum_{\mathbf{q},s=s^{\prime}} v^{\alpha}_{\mathbf{q}, ss^{\prime}}
v^{\beta}_{\mathbf{q},s^{\prime}s} C_{\mathbf{q},s}\tau_{\mathbf{q},s},
\end{equation}
where $\tau_{\mathbf{q},s} = \frac{1}{\Gamma_{\mathbf{q},s}}$, which is the result for the single-mode relaxation-time approximation (SMA) in the Peierls-Boltzmann transport equation (PBTE)~\cite{peierls1997kinetic}.

\begin{figure*}[htp]
	\includegraphics[width = 1.0\linewidth]{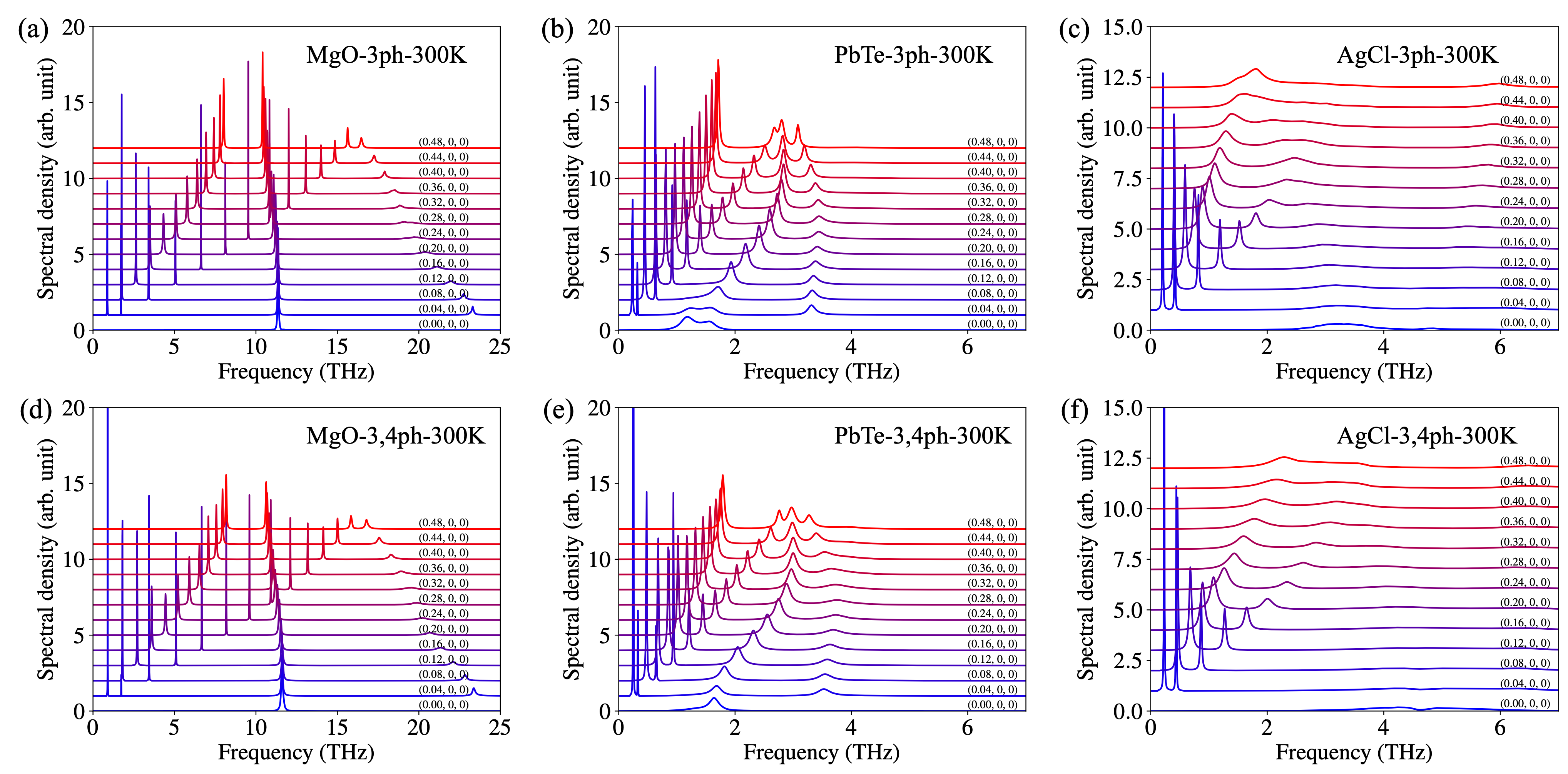}
	\caption{ 
    Calculated phonon spectral density for MgO, PbTe, and AgCl at 300~K along the $\Gamma$–$X$ direction in the first Brillouin zone (fractional coordinates shown in parentheses). Panels (a)–(c) include only three-phonon (3ph) interactions (bubble diagram in Fig.~\ref{fig:diagram}(b)), while panels (d)–(f) additionally include four-phonon (4ph) interactions (sunset diagram in Fig.~\ref{fig:diagram}(c)).
	}
	\label{fig:em_fpsf}
\end{figure*}


\section{Frequency-dependent phonon self-energy}\label{sec:selfeng}
The key ingredient for the calculation of phonon spectral density $b_{\mathbf{q},s}(\omega)$ is the full-frequency dependent self-energy arising from multi-phonon interactions. In this study, we consider 3ph and 4ph interactions from the bubble and sunset diagrams~\cite{balkanski1983anharmonic,xiao2023anharmonic}

\begin{widetext}
\begin{equation}
\label{eq:g3ph}
\begin{split}
\Sigma^{\rm 3ph}_{\mathbf{q},j}(\omega) /\hbar = \Delta^{\rm 3ph}_{\mathbf{q},j}(\omega) - i \gamma^{\rm 3ph}_{\mathbf{q},j}(\omega) =
-\frac{18}{\hbar^2} \sum_{q_1,j_1}\sum_{q_2,j_2} 
\left| V_{\rm 3ph}(q,j; q_1,j_1; q_2,j_2) \right|^2 & \\
\left\{
\frac{n_1+n_2+1}{\omega+i\eta+\omega_1+\omega_2}
-\frac{n_1+n_2+1}{\omega+i\eta-\omega_1-\omega_2}
+\frac{n_1-n_2}{\omega+i\eta-\omega_1+\omega_2}
-\frac{n_1-n_2}{\omega+i\eta+\omega_1-\omega_2}
\right\},
\end{split}
\end{equation}
and
\begin{equation}
\label{eq:g4ph}
\begin{split}
\Sigma^{\rm 4ph}_{\mathbf{q},j}(\omega) /\hbar = \Delta^{\rm 4ph}_{\mathbf{q},j}(\omega) - i \gamma^{\rm 4ph}_{\mathbf{q},j}(\omega) =
-\frac{96}{\hbar^2} \sum_{q_1,j_1}\sum_{q_2,j_2}\sum_{q_3,j_3} 
\left| V_{\rm 4ph}(q,j; q_1,j_1; q_2,j_2; q_3,j_3) \right|^2 & \\
\Biggl\{
\left[ (n_1+1)(n_2+1)(n_3+1)-n_1 n_2 n_3 \right]
\left[ \frac{1}{\omega+i\eta+\omega_1+\omega_2+\omega_3} - \frac{1}{\omega+i\eta-\omega_1-\omega_2-\omega_3} \right]
\\
+3\left[ n_1(n_2+1)(n_3+1)-(n_1+1) n_2 n_3 \right]
\left[ \frac{1}{\omega+i\eta-\omega_1+\omega_2+\omega_3} - \frac{1}{\omega+i\eta+\omega_1-\omega_2-\omega_3} \right]
\Biggl\}
\end{split}
\end{equation}
\end{widetext}
\noindent where $V_{\rm 3ph}$ and $V_{\rm 4ph}$ are the reciprocal cubic and quartic anharmonic coefficients~\cite{balkanski1983anharmonic}, respectively, and $\eta$ is an infinitesimal positive number. The additional frequency-independent loop diagram is included in the self-consistent phonon equation~\cite{tadano2014anharmonic,Errea2011}
\begin{equation}
\label{eq:SCPH1}
\Omega^2_{\lambda} = \omega^2_{\lambda}+2\Omega_{\lambda}\sum_{\lambda_{1}}I_{\lambda\lambda_{1}},
\end{equation}	
where $\omega_{\lambda}$ is the bare frequency of phonon mode $\lambda \equiv (\mathbf{q}, s)$ calculated from the harmonic approximation and $\Omega_{\lambda}$ is the renormalized frequency including temperature effects. The quantity $I_{\lambda\lambda_{1}}$ is defined as
\begin{equation}
\label{eq:SCPH2}
I_{\lambda\lambda_{1}} = \frac{\hbar}{8N} \frac{V^{(4)}(\lambda,-\lambda,\lambda_{1},-\lambda_{1})}{\Omega_{\lambda}\Omega_{\lambda_{1}}} \left[ 1+2n\left(\Omega_{\lambda_{1}}\right) \right],
\end{equation}
where $N$, $\hbar$, $n$, and $V^{(4)}(\lambda,-\lambda,\lambda_{1},-\lambda_{1})$ are respectively the number of sampled wave vectors, the reduced Planck constant, phonon population, and the reciprocal representation of the fourth-order interatomic force constants. The temperature effects in the formalism of SCPH are captured by the phonon population $n$ that obeys the Bose-Einstein statistics. Due to the mutual dependence between $\Omega_{\lambda}$ and $I_{\lambda\lambda_{1}}$, the SCPH equation~\eqref{eq:SCPH1} can be solved in an iterative manner until a reasonable convergence ($10^{-3}$ THz) is reached. 




\section{Computational details}\label{sec:details}
To compute SCPH and full frequency-dependent phonon spectral function, we implemented the above formalisms within the ShengBTE software~\cite{shengbte}. The key ingredients for  in evaluating Eq.(6)-(9) are harmonic and anharmonic force constants. We took advantage of harmonic and anharmonic force constants up to 4th order extracted and validated in our previous studies (MgO from Ref.[\onlinecite{xia2025first}], PbTe from Ref.[\onlinecite{pbte2018}], AgCl from Ref.[\onlinecite{rczbprx}]) using the Compressive Sensing Lattice Dynamics (CSLD) method~\cite{csld,csldlong}. Specifically, we used Perdew-Burke-Ernzerhof (PBE)~\cite{PBE} version of the generalized gradient approximation (GGA)~\cite{gga} for the exchange-correlation (XC) functional~\cite{DFT} for the calculation of MgO, while adopting PBEsol XC functional~\cite{PBEsol} for PbTe and AgCl. Such a choice is based on the report in the literature that PBEsol XC functional can significantly improve the accuracy of predicted lattice parameter and thermodynamic properties of PbTe~\cite{Skelton2014,pbte2018} and AgCl~\cite{rczbprx,ouyang2023role}. We leverage stochastic sampling of scattering events~\cite{guo2024sampling} to accelerate the computation of the full frequency-dependent phonon self-energy.


\begin{figure}[htp]
	\includegraphics[width = 0.85\linewidth]{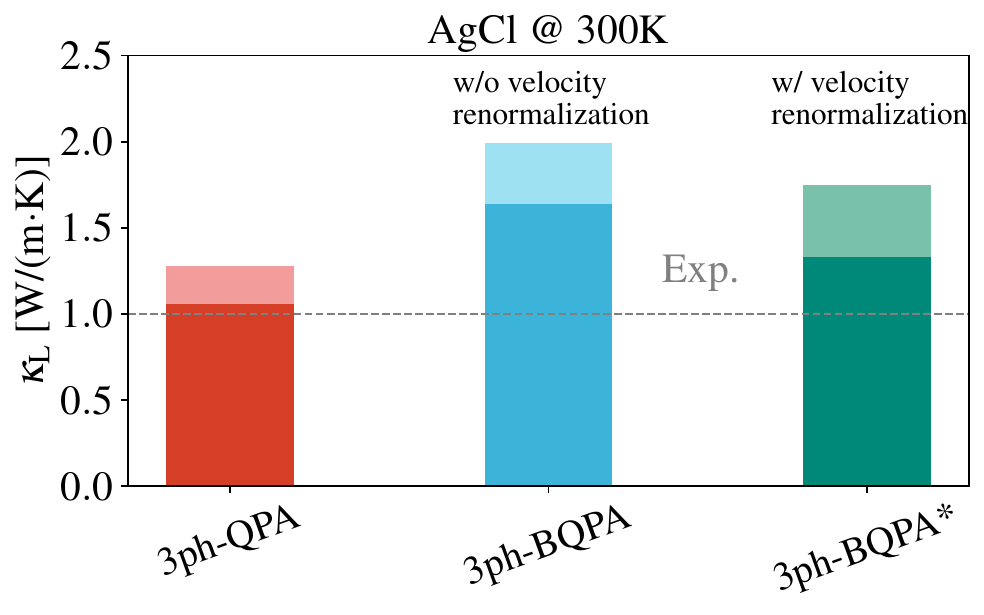}
	\caption{
	Lattice thermal conductivity of AgCl at 300 K comparing three theoretical approaches (3ph-QPA, 3ph-BQPA, and 3ph-BQPA*). The difference between 3ph-BQPA and 3ph-BQPA* arises from the latter further incorporating group-velocity renormalization due to phonon softening. The experimental measurement \cite{maqsood2004thermophysical} is shown as a black dashed line. Dark and light portions of the bars represent the particle-like and coherent contributions to the lattice thermal conductivity, respectively.
	}
	\label{fig:velrenorm}
\end{figure}

\section{Group velocity renormalization}\label{sec:group}

It is worth noting that phonon softening or hardening induced by 3ph or 4ph interactions also modifies the phonon group velocities, which can in turn influence the computed lattice thermal conductivity, $\kappa_{\rm L}$, obtained from Eq.~(\ref{eq:kappa_bqpa}). To assess the significance of this effect, we examine the impact of group-velocity renormalization on $\kappa_{\rm L}^{\rm BQPA}$ for AgCl at 300 K considering only 3ph interactions because this compound exhibits the strongest phonon softening among the three systems studied. Specifically, we identify the quasiparticle peaks in the full phonon spectral functions and perform an inverse Fourier transform of the dynamical matrices at wave vectors commensurate with the supercell to obtain the renormalized force constants. These are then interpolated to evaluate the group velocities throughout the Brillouin zone. As shown in Fig.~\ref{fig:velrenorm}, velocity renormalization leads to a relatively weak reduction in the total $\kappa_{\rm L}$ of AgCl at 300 K, consistent with the phonon softening evident in the spectral functions. Therefore, we see that, for AgCl with only 3ph interactions included, group-velocity renormalization plays a weaker role than the spectral contribution described in Eq.~(\ref{eq:phonon_spectral}). Additional 4ph scattering is not considered here because the phonon spectra in that case are too broadened to reliably identify quasiparticle peaks for the high-lying optical modes. Nevertheless, phonon hardening from 4ph interactions would be expected to slightly increase $\kappa_{\rm L}$.

\end{document}